\documentclass[12pt]{article}

\usepackage{scicite}
\usepackage{bm,amsmath,cases,amssymb,booktabs,mathrsfs,stackrel,graphicx,xcolor,booktabs,hyperref,makecell}
\usepackage{times}
\makeatletter
\renewcommand{\fnum@figure}{\textbf{Fig.~\thefigure}}
\makeatother

\newenvironment{sciabstract}{%
\begin{quote} \bf}
{\end{quote}}

\title{Attentional Ptycho-Tomography (APT) for three-dimensional nanoscale X-ray imaging with minimal data acquisition and computation time}

\author
{Iksung Kang,$^{1\ddagger\dagger\ast}$ Ziling Wu,$^{1\P\dagger}$ Yi Jiang,$^2$ Yudong Yao,$^2$ Junjing Deng,$^2$\\
Jeffrey Klug,$^2$ Stefan Vogt,$^2$ George Barbastathis$^{1,3}$\\
\\
\normalsize{$^{1}$Massachusetts Institute of Technology, Cambridge, Massachusetts 02139, USA}\\
\normalsize{$^{2}$Argonne National Laboratory, Lemont, Illinois 60439, USA}\\
\normalsize{$^{3}$Singapore-MIT Alliance for Research and Technology (SMART) Centre, Singapore}\\
\normalsize{$^{\ddagger}$Present address: University of California, Berkeley, California 94720, USA}\\
\normalsize{$^{\P}$Present address: Singapore-MIT Alliance for Research and Technology Centre, Singapore}\\
\normalsize{$^{\dagger}$These authors contributed equally to this work.}\\
\normalsize{$^\ast$To whom correspondence should be addressed; E-mail: iksung.kang@alum.mit.edu.}
}

\date{}

\begin{document} 


\baselineskip24pt


\maketitle 


\begin{sciabstract}

Noninvasive X-ray imaging of nanoscale three-dimensional objects, e.g. integrated circuits (ICs), generally requires two types of scanning: ptychographic, which is translational and returns estimates of complex electromagnetic field through ICs; and tomographic scanning, which collects complex field projections from multiple angles. Here, we present Attentional Ptycho-Tomography (APT), an approach trained to provide accurate reconstructions of ICs despite incomplete measurements, using a dramatically reduced amount of angular scanning. Training process includes regularizing priors based on typical IC patterns and the physics of X-ray propagation. We demonstrate that APT with 12-time reduced angles achieves fidelity comparable to the gold standard with the original set of angles. With the same set of reduced angles, APT also outperforms baseline reconstruction methods. In our experiments, APT achieves 108-time aggregate reduction in data acquisition and computation without compromising quality. We expect our physics-assisted machine learning framework could also be applied to other branches of nanoscale imaging.
\end{sciabstract}


\section*{Introduction}

Three-dimensional X-ray imaging enables noninvasive monitoring of objects' interiors with nanoscale resolution. Integrated circuits (IC) are especially interesting for this operation, for two reasons: first, noninvasive inspection of ICs is important for verifying manufacturing integrity. Second, ICs follow specific design rules, which makes their geometries highly regular and yet highly diverse. The geometrical properties are then useful as prior knowledge, enabling vast improvements in practical aspects of the imaging process, such as acquisition time as we show here.

Prior works have typically used two types of scanning: translational and rotational. The translational scan ({\em ptycho}) is inspired by ptychography, \textit{i.e.} a scanning-based coherent diffraction imaging method for phase retrieval. Ptychography was originally proposed by W. Hoppe~\cite{hoppe1969beugung} to solve the phase problem in Scanning Transmission Electron Microscopy (STEM), where a moving aperture resolves the ambiguity in phase based on translational invariance. The term ``ptychography'' was coined in the following year~\cite{hegerl1970dynamische}. Nellist {\it et al.}~\cite{nellist1995resolution} demonstrated resolution improvement in STEM by a factor of $2.5$ over the limit imposed by partial coherence, exploiting the redundancy in the ptychographic measurements. As an alternative that does not even require careful aberration correction in the optics, Gerchberg and Saxton~\cite{gerchberg1972practical} introduced a lensless iterative phase retrieval algorithm, now referred to as \textit{GS} after them. This work was extended to lensless ptychography for extended objects by Faulkner~\cite{faulkner2004movable}. Subsequently, Rodenburg~\cite{rodenburg2004phase} introduced yet another iterative phase retrieval algorithm called Ptychographical Iterative Engine (PIE) that simultaneously retrieves both the object and the probe function. Thus, the requirement of a high-quality lens for imaging is fundamentally lifted. Further advances by Thibault {\it et al.}~\cite{thibault2008high} and Thibault and Guizar-Sicairos~\cite{thibault2012maximum} led to the Difference Map (DM) algorithm and Maximum Likelihood algorithm, respectively, for iterative ptychographic reconstruction.

After the ptychographic reconstruction step, the second angular scan ({\em tomo}) operation is required to retrieve the object's interior, as in tomography. For parallel beam illumination and under the weak scattering approximation, the measurements are interpreted simply as projections through the object, {\it i.e.} the measurements implement the object's Radon transform~\cite{radon1917determination,radon1986determination}. The inverse Radon transform is typically implemented as a version of the Filtered Back-Projection (FBP) algorithm, first proposed by Bracewell and Riddle~\cite{bracewell1956strip,bracewell1967inversion}. Gordon, Bender, and Herman~\cite{gordon1970algebraic} proposed an alternative iterative tomography algorithm called Algebraic Reconstruction Technique (ART) that iteratively, which applies also to non-parallel illumination beams and works by updating the object estimate to sequentially bring each reconstructed projection into agreement with the corresponding measured projection. Subsequent improvements of this original iterative method were the Simultaneous Iterative Reconstruction Technique (SIRT)~\cite{gilbert1972iterative} and the Simultaneous Algebraic Reconstruction Technique (SART)~\cite{andersen1984simultaneous}, which consider all projections simultaneously and thus drastically reduce the number of iterations for the reconstruction process. Maximum Likelihood methods have also been popular for tomography, with the Bouman-Sauer algorithm~\cite{bouman1993generalized} as one of the most prominent.

For X-rays, the high penetration depth facilitates recovery of information deep inside the sample in the angular sampling scheme. Combining this property with translational scanning for lensless high spatial resolution, Dierolf {\it et al.}~\cite{dierolf2010ptychographic} proposed the Ptychographic X-ray Computed Tomography (PXCT) scheme to determine the volumetric interior of biological specimens with nanoscale details. Using this technique, Holler {\it et al.}~\cite{holler2017high} experimentally demonstrated noninvasive imaging of ICs produced with $22\:\text{nm}$ technology at $14.6\:\text{nm}$ resolution. These techniques are limited by the requirement for two types of scanning, angular and translational, and scale badly scales with object volume. A novel X-ray microscope called Velociprobe~\cite{deng2019velociprobe} utilizes fly-scan ptychography~\cite{huang2015fly} to significantly reduce the data acquisition time. Still, total data acquisition and reconstruction time for a typical $100\times 100\times 5\:\mu\text{m}^3$ IC $(\simeq 2\times 10^{10}\:\text{voxels})$ is estimated to be in excess of two months.

Here, we propose a machine learning framework to reduce data acquisition and computation time for IC reconstruction under the X-ray ptycho-tomography geometry. The reduction in data acquisition is compensated by explicit use of prior knowledge of the typical objects being imaged, and of the optical physics of the imaging system. The length of the acquisition and computation time scale as the number $N$ of tomo-scans. The total angular range $\theta$ determines the size of the missing wedge in the Fourier domain and, therefore, is commensurate with loss of fidelity. Our ``gold standard'' is a ptycho-tomo reconstruction by SART with $N = 349$ and $\theta = \pm 70.4^\circ$. This maximum angle is determined by practical considerations, such as the sample geometry. More details about the gold standard geometry and our approach are available in Methods.

\begin{figure}[htbp!]
    \centering
    \includegraphics[width=\linewidth]{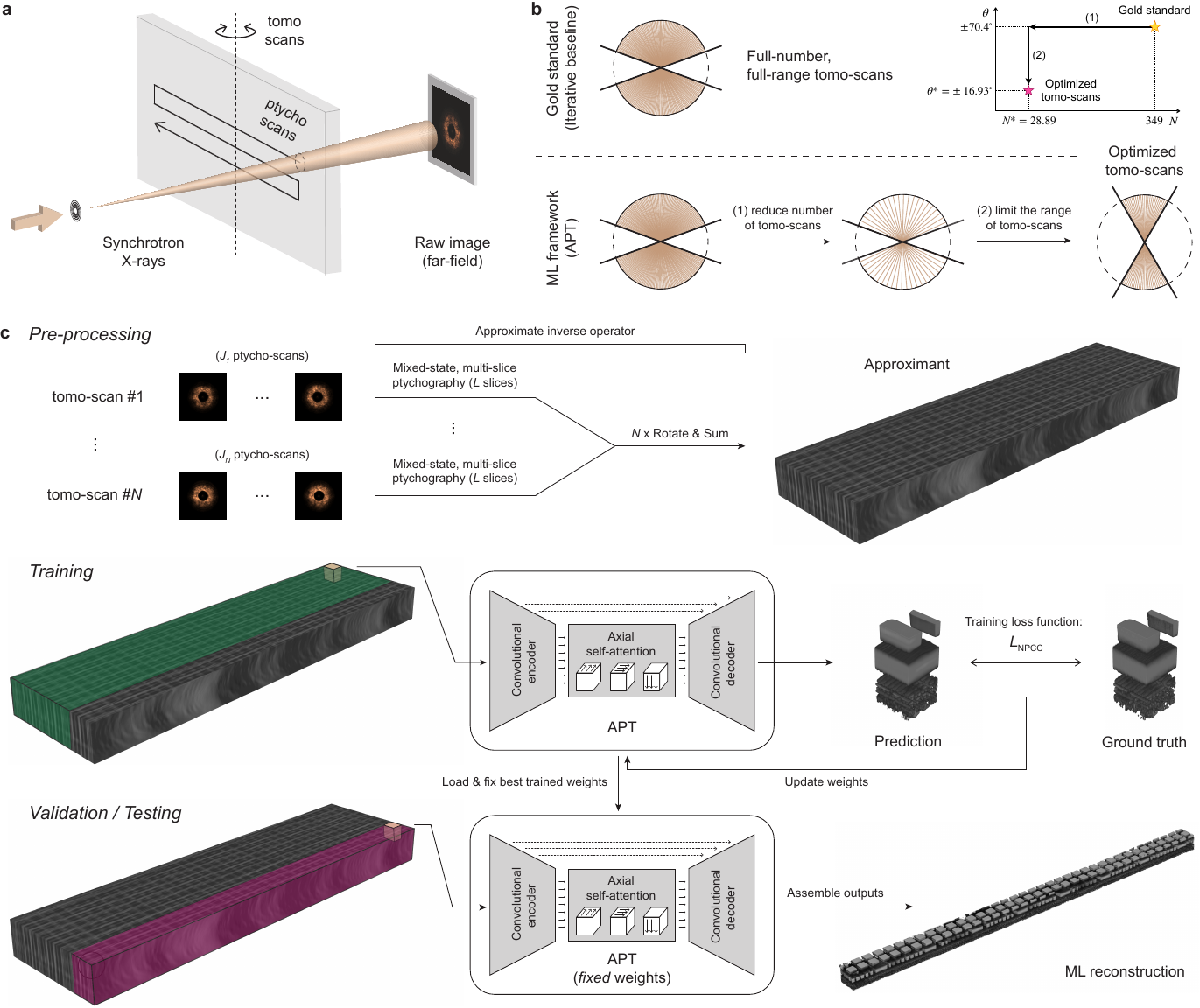}
    \caption{\textbf{X-ray ptycho-tomography and the implementation of APT.} \textbf{(a)} Brief schematic of X-ray ptycho-tomography geometry with translational scanning of synchrotron X-rays (\textit{ptycho-scans}) and symmetric angular scanning of the IC sample with uniform angular increment (\textit{tomo-scans}). \textbf{(b)} Gold standard uses $349$ tomo-scans within the angular range of $\pm 70.4^\circ$, but our machine learning framework (APT) uses fewer tomo-scans optimized through two steps. \textbf{(c)} Diffraction intensities are pre-processed with an approximate inverse operator to generate the Approximant (and more details can be found in Methods and Supplementary Materials.) One of two non-overlapping portions of the Approximant is used for training with a negative Pearson correlation coefficient (NPCC) as the training loss function, where network weights are updated over several training epochs. For testing, best trained weights are loaded and fixed to generate outputs over the test volume ($4.48\times 93.2\times 3.92\:\mu\text{m}^3$).}
    \label{fig:introduction}
\end{figure}

To search this two-dimensional space $(N,\theta)$, our strategy is as follows: we start with the gold standard nominal values of $N$ and $\theta$. If we reduced $N$ while using a standard reconstruction algorithm, like FBP, SIRT, etc. mentioned earlier, performance would decrease immediately. With machine learning, we find that it is possible to regularize for the loss of angular sampling density and still maintain reconstruction fidelity, down to a minimum number $N^*$. Then we start reducing the total angular range, meaning that the sampling now becomes denser. The machine-learning regularizer again manages to maintain approximately even fidelity down to a minimum range $\theta^*$. This is our optimal operating point $(N^*, \theta^*).$ The strategy is depicted in Fig.~\ref{fig:introduction}b. In principle, this procedure can be repeated to find even tighter operating conditions, but we did not carry that out as we would expect any further gains to be minimal.

That machine learning is suitable for achieving even fidelity while the amount of sampled data is decreasing is not entirely surprising by now. The key is the ability of deep neural networks to very effectively capture regularizing priors, especially sparsity, in both supervised mode as we do here and in untrained mode~\cite{ulyanov2018deep,mataev2019deepred}. Previous demonstrations of supervised learning have been carried out for Fourier ptychography~\cite{kappeler2017ptychnet,nguyen2018deep} and two-dimensional ptychography~\cite{cherukara2020ai}. The reason we chose the supervised learning mode is because we had ample data available from the gold standard ptycho-tomo approach. 

APT is described schematically in Fig.~\ref{fig:introduction}c. We first invert the far-field diffraction intensities (or ptycho-scans) with an approximate inversion operator. This yields to get an approximate volumetric estimate of the interior of an IC chip, which we dub the ``Approximant''~\cite{goy2018low}. This step utilizes prior knowledge on underlying imaging physics and pre-processes the input with the physics prior. The Approximant as a result of this pre-processing step (or physics-informing step) is defective in terms that layers are not well separated because of the approximate inversion from diffraction intensities and only a small fraction of tomo-scans used for the computation. During training, the neural network's weights are optimized based on the Approximant as input. Upon the completion, the trained neural network gives a refined volumetric reconstruction of ICs.

The proposed neural network is based on a 3D U-net architecture~\cite{ronneberger2015u,cciccek20163d} and augmented with the multi-head axial self-attention~\cite{wang2020axial} to address lack of spatial resolution in the Approximant by taking advantage of its global-range interactions to retrieve information from all layers to resolve each layer's structure. We choose this multi-head axial self-attention over multi-head self-attention~\cite{vaswani2017attention} to alleviate computational burden.

We demonstrate that the present method is capable of providing reliable reconstructions of ICs even when both the number and the total angular range of tomo-scans are largely decreased to $N^* \sim 29$ and $\theta^* \sim \pm 17^\circ$ representing an improvement of $\times 12$ and $\times 4.2$, respectively. For the reconstruction of an IC chip over the test volume ($4.48\times 93.2\times 3.92\:\mu\text{m}^3$), $0.63$ hours (or $38$ minutes) is sufficient for both data acquisition and reconstruction with our machine learning framework. The improvements work out to an approximate overall $\times 108$ reduction in total (acquisition plus computation) time compared to the current up-to-date iterative reconstruction method.

\section*{Results}
\subsection*{Reducing acquisition and scanning time \label{sec:reducing}}

The synchrotron beam is delivered on the sample, and a full lateral scan is carried out to obtain the ptychographic information for each angular orientation of the sample. Repeating for $N$ angles collects tomographic information for the interior's reconstruction. The raw intensities past the sample are recorded by a digital camera detector at each scan position. The details of the experimental collection system are in Methods. The collected raw intensities are then processed in two steps: the first step embeds the physics of X-ray propagation through an Approximant operator~\cite{goy2019high,kang2021dynamical}, while the second step consists of the APT network delivering the final reconstruction, as described earlier. The details of training and operating this computational pipeline are in Methods. As discussed earlier, our approach is to first reduce scanning time by finding the minimum $N^*$ and then reduce computation time by finding the minimum $\theta^*$. 

\begin{figure}[htbp!]
    \centering
    \includegraphics[width=\linewidth]{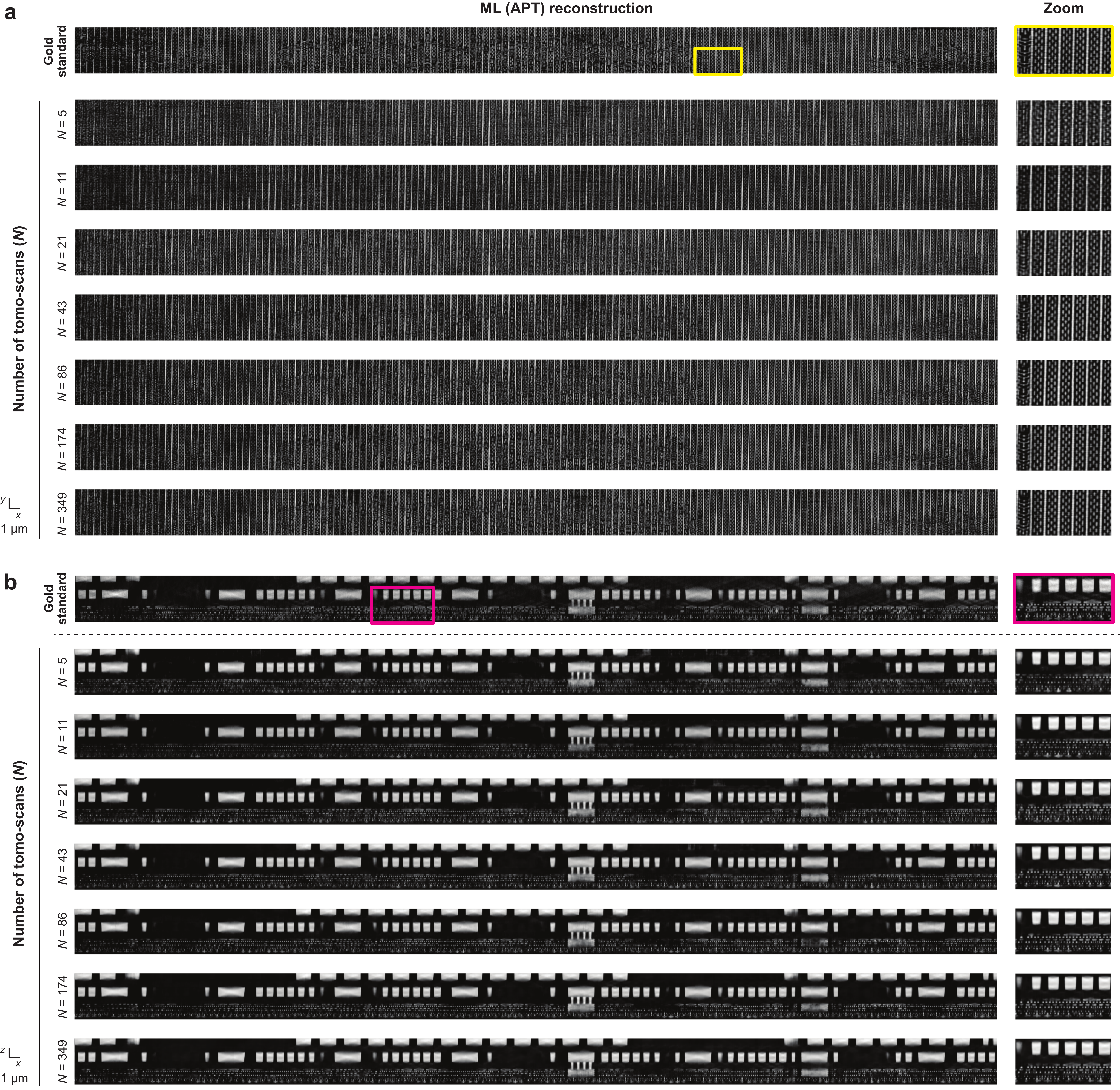}
    \caption{\textbf{Optimizing the number of tomo-scans - qualitative view.} Qualitative comparison from a parameter sweep over the number of tomo-scans ($N$) at two different depths: \textbf{(a)} $z = 0.364\:\mu\text{m}$ and \textbf{(b)} $y = 0.532\:\mu\text{m}$. The figure shows APT reconstructions with different $N$ over the test volume ($4.48\times 93.2\times 3.92\:\mu\text{m}^3$).}
    \label{fig:qualitative_comparison_ds}
\end{figure}

\begin{figure}[htbp!]
    \centering
    \includegraphics[width=\linewidth]{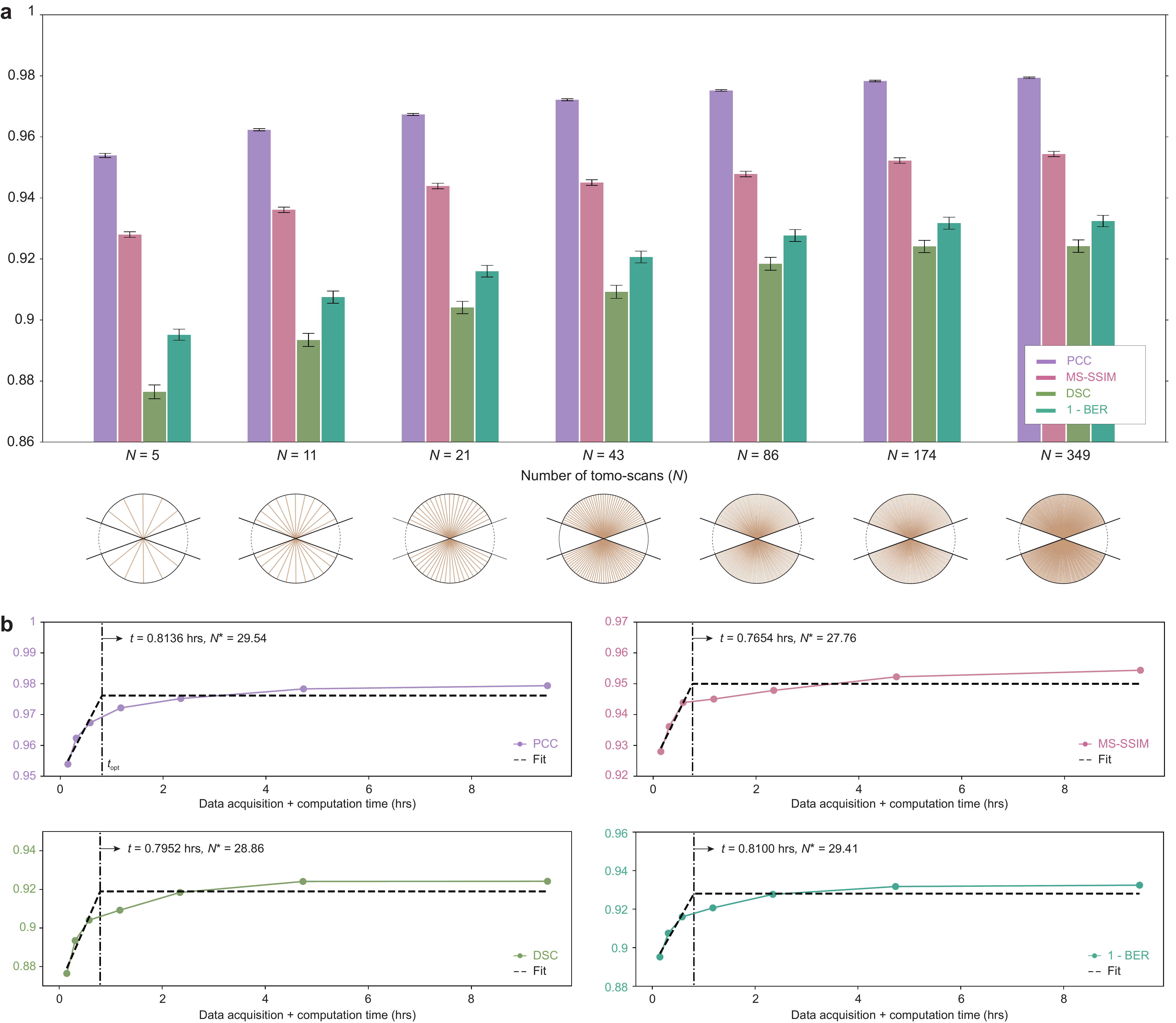}
    \caption{\textbf{Optimizing the number of tomo-scans - quantitative view.} \textbf{(a)} Quantitative comparison from a parameter sweep over the number of tomo-scans ($N$) with four different quantitative metrics. \textbf{(b)} The number of tomo-scans that optimally compromise the performance ($N^*$) is $28.89$ in average, where APT reduces the data acquisition and computation time by a factor of $85$.}
    \label{fig:ds_bar_threshold}
\end{figure}

A parameter sweep over $N$ is shown qualitatively in Fig.~\ref{fig:qualitative_comparison_ds}. Four quantitative performance comparisons are shown in Fig.~\ref{fig:ds_bar_threshold}, in terms of the following metrics: Pearson correlation coefficient (PCC)~\cite{benesty2009pearson}, multi-scale structural similarity index metric (MS-SSIM)~\cite{wang2003multiscale}, Dice Similarity coefficient (DSC)~\cite{dice1945measures}, and Bit-Error Rate (BER, more details available in Methods). Both analyses indicate that $N^*\sim 29$, representing a reduction of more than $\times 12$ over the gold standard of $N = 349$. Reducing $N$ significantly below this value results in noticeable degradation, both qualitatively and quantitatively.

\begin{figure}[htbp!]
    \centering
    \includegraphics[width=\linewidth]{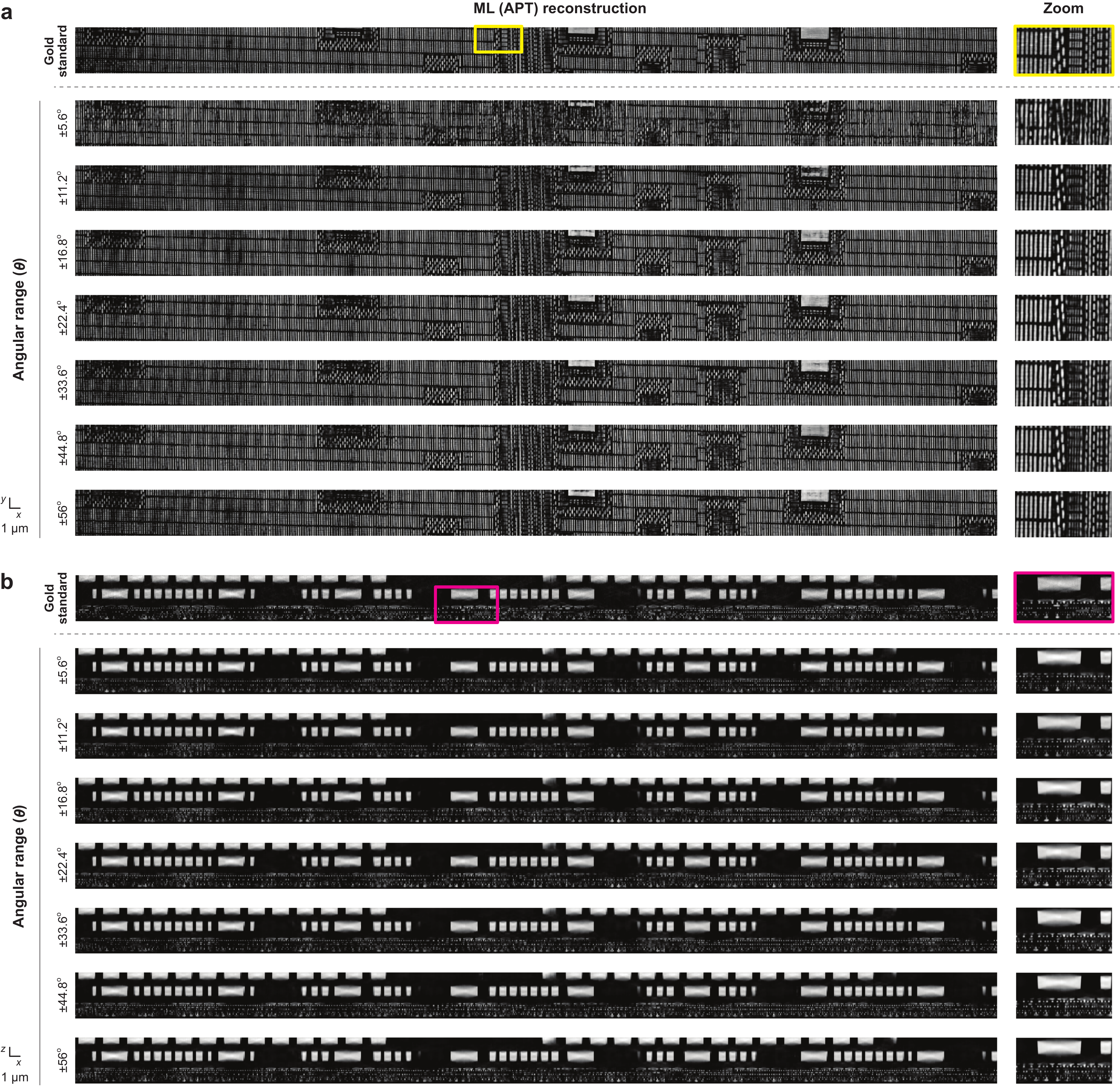}
    \caption{\textbf{Optimizing the number of angular range - qualitative view.} Qualitative comparison from a parameter sweep over the angular scanning range at two different depths: \textbf{(a)} $z = 1.092\:\mu\text{m}$ and \textbf{(b)} $y=4.186\:\mu\text{m}$. The figure shows APT reconstructions over the test volume ($4.48\times 93.2\times 3.92\:\mu\text{m}^3$).}
    \label{fig:qualitative_comparison_pms}
\end{figure}

\begin{figure}[htbp!]
    \centering
    \includegraphics[width = \linewidth]{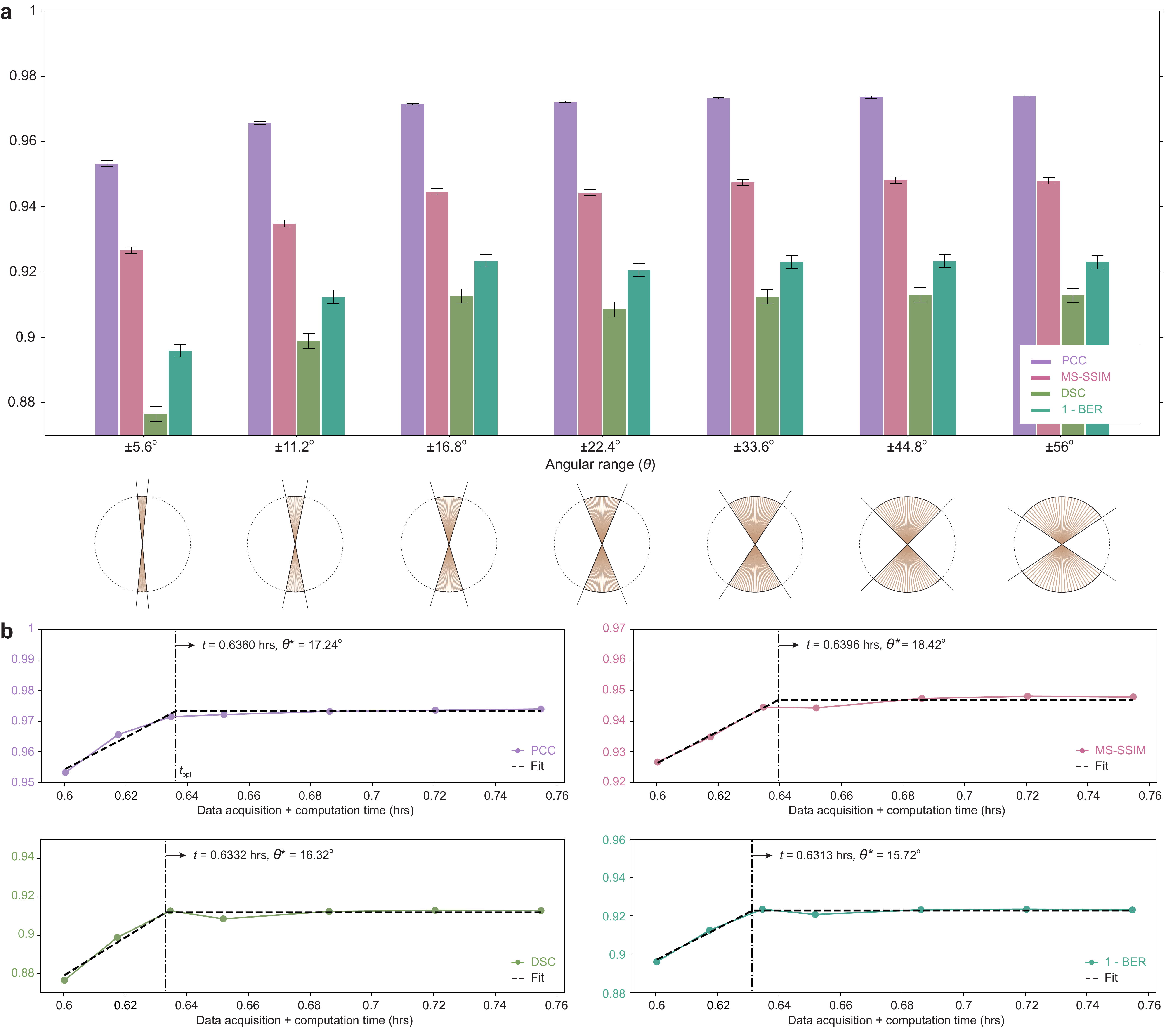}
    \caption{\textbf{Optimizing number of angular range - quantitative view.} \textbf{(a)} Quantitative comparison from a parameter sweep over the angular scanning range at $N = N^*\:(= 29)$. \textbf{(b)} The total angular range that optimally compromise the performance ($\theta^*$) is $\theta^* = \pm 16.93^\circ$ in average. APT decreases the time for whole process by $108$ times.}
    \label{fig:pms_bar_threshold}
\end{figure}

Next we fix $N=29$ and perform a parameter sweep over $\theta$. Qualitative results are shown in Fig.~\ref{fig:qualitative_comparison_pms}, while the quantitative evaluation according to the same four metrics of the previous section is in Fig.~\ref{fig:pms_bar_threshold}. Both analyses lead to $\theta^*\sim \pm 17^\circ$ as the approximate lower bound before drastic degradation occurs. The savings in data acquisition and computation times are $\times 12$ and $\times 105$, respectively, and total time savings (acquisition plus computation) of $\times 108$. 

\subsection*{Regularization and imaging system physics \label{sec:regphy}}

\begin{figure}[htbp!]
    \centering
    \includegraphics[width = \linewidth]{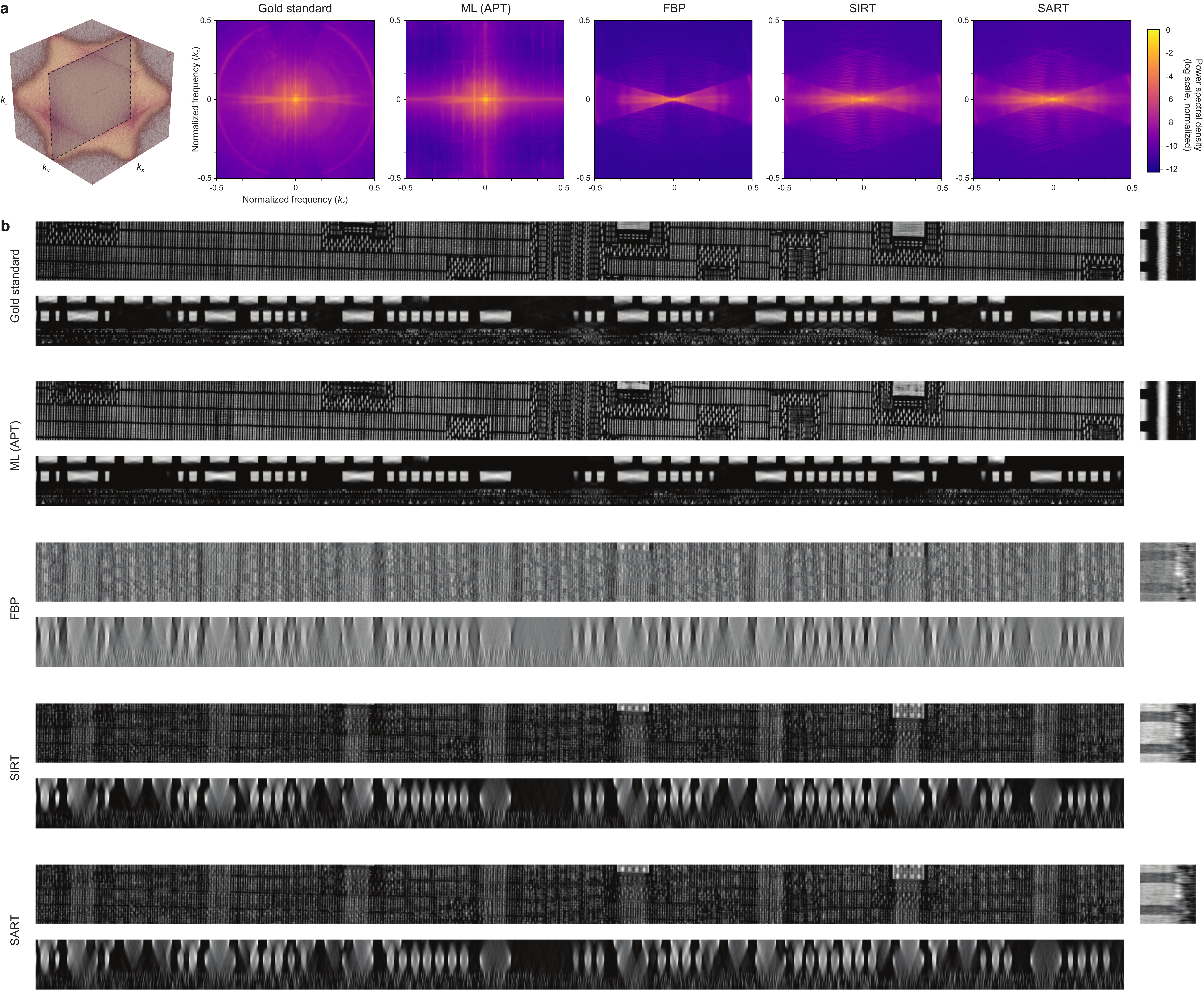}
    \caption{\textbf{Reconstruction performance comparison with baseline methods.} \textbf{(a)} The figure qualitatively compares 3D power spectral densities of reconstructions of gold standard and baseline reconstruction methods. Both APT and baseline reconstruction methods (FBP, SIRT, SART) use the optimal tomo-scans with $N^*= 29$ and $\theta^*= \pm 16.8^\circ$. Only APT has effectively filled up the missing wedges. \textbf{(b)} Reconstructions of gold standard and baseline methods are shown and compared along $xy$, $xz$, and $yz$ planes.}
    \label{fig:psd_reconstruction_comparison}
\end{figure}

The reported improvements suggest that the APT algorithm is particularly effective at learning regularizing priors to compensate for the missing information. Fig.~\ref{fig:psd_reconstruction_comparison}a shows the power spectral densities of the gold standard, the APT reconstruction, and the baseline tomographic reconstruction methods FBP~\cite{bracewell1956strip}, SIRT~\cite{gilbert1972iterative}, and SART~\cite{andersen1984simultaneous}, all obtained at $N^*=29$ and $\theta^*=\pm 16.8^\circ$. The missing wedge is evident in the latter three. The qualitative cross-sections in Fig.~\ref{fig:psd_reconstruction_comparison}b confirm that the missing wedge effect leads to severe artifacts in the baseline methods, but not in APT. 

APT also relies on its input, the Approximant, having carefully taken into account the physics of the imaging system. Unlike earlier works where the illumination on the sample was coherent~\cite{goy2019high,kang2021dynamical}, the synchrotron may be considered as temporally coherent but is rather less coherent spatially. The mutual intensity is expressed as a linear combination of mutually incoherent states, also known as coherent modes~\cite{wolf1982new}. Accounting correctly for the synchrotron X-ray's coherence state has been shown to improve spatial resolution and phase contrast in standard ptychography for thin samples~\cite{thibault2013reconstructing}. 

For samples thicker than the depth of focus of the probe, multi-slice reconstruction from simple ptychography has been demonstrated with visible light~\cite{tsai2016x}, X-rays~\cite{suzuki2014high} and electrons~\cite{Chen826}. This is the starting point for our Approximant (please see Fig.~\ref{fig:introduction}a.) We form the cost function 
 \begin{equation}\label{eq:L}
    \mathcal{L}_n = \sum_{j=1}^{J_n}\sum_{m=1}^M\sum_\mathbf{q} \left(\left|\mathcal{F}\left\{P_{j,\mathbf{r},m}^{(n)[L]}O_\mathbf{r}^{(n)[L]}\right\}\right| - \sqrt{I_{j,\mathbf{q}}^{(n)}}\right)^2\:\:\:(n=1,2,\cdots,N),
\end{equation}
where $N$ is the number of given tomo-scans, $J_n$ the number of ptycho-scans associated with the $n$-th tomo-scan; $M$ the number of coherent modes; $L$ the number of slices for our given depth of focus works out to equal to $5$; $\mathbf{q}$ denotes the coordinates in the reciprocal space; and $P_{j,\mathbf{r},m}^{(n)[L]}$, $I_{j,\mathbf{q}}^{(n)}$ indicate the wavefield before the $L$-th slice from the $m$-th coherent mode and experimental diffraction intensity at the $j$-th ptycho- and the $n$-th tomo-scans, respectively. We run two iterations of a gradient scheme on Eq.~\ref{eq:L} and obtain the argument $\angle O^{(n)[l]}_{\mathbf{r}}$ at each one among the $l=1,\cdots,\:L$ slices~\cite{tsai2016x,Chen826}. We rotate the result back to the original coordinate system, and average the estimates from all tomo-scan steps to yield the final Approximant. More details can be found in Methods. 

The Approximant computation step is the slowest in the pipeline; in our computing hardware (see Methods), it takes $36$ minutes when $\theta = \pm 70.4^\circ$ and $26$ minutes when $\theta\sim\pm 17^\circ$. In addition to the computation time, the spacing between slices in the Approximant is nominally limited by the depth of focus, and that is why we only reconstruct $L=5$ of them. The number of desired reconstruction slices is much larger, \textit{i.e.} 280, so we simply dilate the Approximant slices to match it. As a result, the input to the neural network is poor (more in Supplementary Materials). Nevertheless, the subsequent APT architecture learns how to use the multi-slices as input and, as long as $N>N^*$ and $\theta>\theta^*$, produce a high-fidelity final reconstruction with much finer slice spacing.

\section*{Discussion}

APT is trained using the gold standard reconstructions of randomly selected segments from a single IC specimen, which was made available for our experiments. This prompts us to address two related concerns: {\it (1)} what can we guarantee about fidelity of the gold standard and, hence, our reconstructions {\it vis-\`a-vis} the ground truth, {\it i.e.} the physical specimen? and {\it (2)} is our APT overtrained to this specific IC? 

The first concern was partially addressed by Refs.~\cite{goy2019high,kang2021dynamical}, where the design files of the geometrical features were treated as ground truth. (That method was still bound by the assumption that the physical specimen matched the design files; but that was less of a concern, given the size scales involved.) Neither of these algorithms would have worked in the case reported here, because of the great range of feature sizes in the specimen and because the synchrotron X-ray is not spatially coherent. Moreover, the design files for the specimen are not available to us. On the other hand, the gold standard was obtained quite thoroughly with $95\%$ spatial overlap factor in the ptycho-scan and $N=349$ angles in the tomo-scan. Besides, there are no discernible visual artifacts in the gold standard reconstructions. These facts provide us with reasonable assurance about the fairness of our comparisons in Figs.~\ref{fig:qualitative_comparison_ds}-\ref{fig:psd_reconstruction_comparison}. 

Regarding the second concern: if new structures are given where the priors are {\em significantly} different than the priors learnt here (e.g. features oriented at $45^\circ$) then APT would have to be retrained. This is a necessary limitation of our supervised learning approach. The same holds for non-IC objects such as viruses. If, moreover, not enough physical specimens are available for supervised training, then it is possible to train by rigorously simulating the forward propagation of X-rays through the specimen (as Refs.~\cite{goy2019high,kang2021dynamical} did for visible light) or use ``untrained'' methods, such as deep image prior~\cite{ulyanov2018deep,baguer2020computed}.

The reported best values of $N^*\sim 29$ and $\theta^\star\sim\pm 17^\circ$ are not fundamental, but indicative of the effectiveness of IC geometries acting as regularizing prior. Less complex geometries, smooth and with less content at high frequencies in the missing wedge, could achieve even better reductions, whereas complex structures with smaller features and higher refractive index contrast would be more limited. A full theoretical analysis of how $N^*$ and $\theta^*$ depend on the complexity of the prior is beyond the scope of this work.

Lastly, regarding ICs in particular and planar samples more generally, the total attenuation of the X-rays increases at large angles, which leads to artifacts. It may be compensated computationally, or by scanning the illumination wavevectors on a conical surface. The latter scheme is referred to as laminography~\cite{helfen2005high,holler2019three}. It is beyond the scope of our present work, but it would be interesting to investigate if approaches similar to the one reported here are applicable.

\section*{Methods}
\subsection*{Experiment and the gold standard preparation}\label{sec:ground_truth}
Integrated circuits produced with $16\mbox{-}\text{nm}$ technology of size $25.1\times 93.2\times 3.92\:\mu\text{m}^3$ were laterally scanned with synchrotron X-rays of $8.8\:\text{keV}$ for each tomo scan with Velociprobe~\cite{deng2019velociprobe} at the Advanced Photon Source (APS) of the Argonne National Laboratory (ANL). $12$ coherent modes of the synchrotron X-ray were used for the experiment. Tomo-scans were carried out from $\mbox{-}70.4^\circ$ to $70.4^\circ$ with angular increment of $0.4^\circ$, and for each tomo-scan, ptycho-scans were recorded on-the-fly at $\sim 60$k lateral positions with Dectris Eiger X 500K area detector (pixel size: $75\:\mu\text{m}$, sample-to-detector distance: $1.92\:\text{m}$) at a frame rate of $500\:\text{Hz}$. Elapsed time of this whole data acquisition process (translational and angular) was $12.51\:\text{hrs}$, or $129$ seconds per tomo-scan.

As a first step to obtain the gold standard reconstruction, a two-dimensional projection was reconstructed for each tomo-scan with $600$ iterations of the least-square maximum likelihood ptychographic algorithm~\cite{odstrvcil2018iterative} as implemented in PtychoShelves~\cite{wakonig2020ptychoshelves}. Raw diffraction intensities of $256\times 256\:\text{px}^2$ size were downsampled by $\times 2$ to accelerate the computation. The ptychographic reconstruction for all $349$ tomo-scans was processed with $8$ Tesla V$100$ GPUs in parallel to expedite the process, thus taking $362.09\:\text{hrs}$ for this step.

Then, the projections were aligned to a tomographic rotation axis with an additional correction in the form of a phase ramp removal process. Then, a deep neural network pre-trained on similar images of integrated circuits was applied to the aligned projections for upsampling by $\times 2$~\cite{dong2014learning,dong2015image}. The elapsed time of this step was approximately $5\:\text{hrs}$. 

Lastly, the final tomographic reconstruction was performed using $349$ upsampled projections with $10$ iterations of SART to generate a finally three-dimensional reconstruction of the IC sample with the isotropic $14\mbox{-}\text{nm}$ voxel size, which took $1\:\text{hr}$ with $8$ Tesla V$100$ GPUs.

\subsection*{Gradient calculation}\label{sec:gradient}

Considering the mixed-state (spatially partially coherent) nature of synchrotron X-rays and multi-slice structure of the IC sample, a forward model can be formulated as
\begin{gather}\label{eq:forward_model}
    \psi_{j,\mathbf{r},m}^{(n)[L]} = O_\mathbf{r}^{(n)[L]}\mathcal{P}_{\Delta z}\left[O_\mathbf{r}^{(n)[L-1]}\mathcal{P}_{\Delta z}\left[\cdots\mathcal{P}_{\Delta z}\left[P_{\mathbf{r\mbox{-}r_j},m} O_\mathbf{r}^{(n)[1]}\right]\right]\right],\\
    I_{j,\mathbf{q}}^{(n)} = \sum_{m=1}^M\left|\Tilde{\psi}_{j,\mathbf{q},m}^{(n)[L]}\right|^2 = \sum_{m=1}^M\left|\mathcal{F}\left[\psi_{j,\mathbf{r},m}^{(n)[L]}\right]\right|^2,
\end{gather}
where 
\begin{itemize}
    \item[\mbox{-}] $n$: the index of tomo-scans ($n=1,2,\cdots, N$)
    \item[\mbox{-}] $j$: the index of ptycho-scans ($j=1,2,\cdots, J_n$)
    \item[\mbox{-}] $l$: the index of multi-slices ($l=1,2,\cdots, L$)
    \item[\mbox{-}] $m$: the index of mixed-states ($m=1,2,\cdots, M$)
    \item[\mbox{-}] $\mathbf{r}$: the spatial domain coordinates
    \item[\mbox{-}] $\mathbf{q}$: the reciprocal domain coordinates
    \item[\mbox{-}] $P_{\mathbf{r},m}$: the $m$-th coherent mode of the synchrotron X-ray probe
    \item[\mbox{-}] $O_\mathbf{r}^{(n)[l]}$: the $l$-th slice of the object viewed at the $n$-th tomo-scan. 
\end{itemize}

The following describes the gradient computation of the loss function in Eq.~\ref{eq:L} based on the forward model, which was done automatically with Ptychoshelves~\cite{wakonig2020ptychoshelves}. The gradients of the loss function with respect to the wavefield and the complex object are
\begin{subnumcases}{\frac{\partial\mathcal{L}}{\partial P_{j,\mathbf{r},m}^{(n)[l]}} = }
    O_\mathbf{r}^{(n)[l]} \left(\mathcal{P}_{-\Delta z}\left\{\frac{\partial\mathcal{L}}{\partial P_{j,\mathbf{r},m}^{(n)[l+1]*}}\right\}\right)^* & \text{for} $1\leq l < L$\label{eq:gradient_probe_1}\\
    O_\mathbf{r}^{(n)[L]} \chi_{j,\mathbf{r},m}^{(n)[L]*} & \text{for} $l = L$\label{eq:gradient_probe_2},
\end{subnumcases}

\begin{subnumcases}{\mathrm{and}\quad\frac{\partial\mathcal{L}}{\partial O_r^{(n)[l]}} = }
    \sum_{j=1}^{J_n} \sum_{m=1}^M P_{j,\mathbf{r},m}^{(n)[l]}\left(\mathcal{P}_{-\Delta z}\left\{\frac{\partial\mathcal{L}}{\partial P_{j,\mathbf{r},m}^{(n)[l+1]*}}\right\}\right)^* & \text{for} $1\leq l < L$\label{eq:gradient_object_1}\\
    \sum_{j=1}^{J_n} \sum_{m=1}^M P_{j,\mathbf{r},m}^{(n)[L]}\chi_{j,\mathbf{r},m}^{(n)[L]*} & \text{for} $l = L$\label{eq:gradient_object_2},
\end{subnumcases}

\begin{gather}
    \mathrm{where}\quad P_{j,\mathbf{r},m}^{(n)[l]} = \mathcal{P}_{\Delta z}\left[O_\mathbf{r}^{(n)[l-1]}P_{j,\mathbf{r},m}^{(n)[l-1]}\right],\\
    \frac{\partial\mathcal{L}}{\partial P_{\mathbf{r},m}} = \sum_{n=1}^N\sum_{j=1}^{J_n}\frac{\partial\mathcal{L}}{\partial P_{j,\mathbf{r}+\mathbf{r}_j,m}^{(n)[1]}},\\
    \chi_{j,\mathbf{r},m}^{(n)[L]} = \mathcal{F}^{-1}\left\{\left(1 - \frac{\sqrt{I_{j,\mathbf{q}}^{(n)}}}{\left|\Tilde{\psi}_{j,\mathbf{q},m}^{(n)[L]}\right|}\right)\Tilde{\psi}_{j,\mathbf{q},m}^{(n)[L]}\right\}.
\end{gather}

With two iterations of gradient descent on the loss function in Eq.~\ref{eq:L}, we obtain the multi-slice estimate $O_\mathbf{r}^{(n)[l]}\rvert_{N_{\text{iter}}=2}$ for each tomo-scan and subsequently its argument at each one of the $L=5$ slices. For the final Approximant, we rotate the results back to the original coordinate system, and average $N$ estimations from all $N$ tomo-scans. Please see Supplementary Materials for visualization. More details on the gradient calculation can be found in Refs.~\cite{tsai2016x,Chen826}.

\subsection*{Machine learning framework}\label{sec:ml_framework}

Our neural network architecture is based on a 3D U-net structure~\cite{ronneberger2015u,cciccek20163d} augmented with multi-head axial self-attention (``axial self-attention'' in short)~\cite{wang2020axial}. The U-net directly transfers multi-scale features to its decoder arm to preserve spatial information, and the axial attention augments the features with its global-range self-interactions. 

The U-net backbone encoder design was influenced by the well-established architecture ResNet50~\cite{he2016deep} with some modifications so that it can accommodate 3D instead of 2D data. The architecture's decoder then upsamples the features by $\times 2$ to result in isotropic voxels of linear size $14\:\text{nm}$. More details can be found in Supplementary Materials.

\begin{figure}[htbp!]
    \centering
    \includegraphics[width=\linewidth]{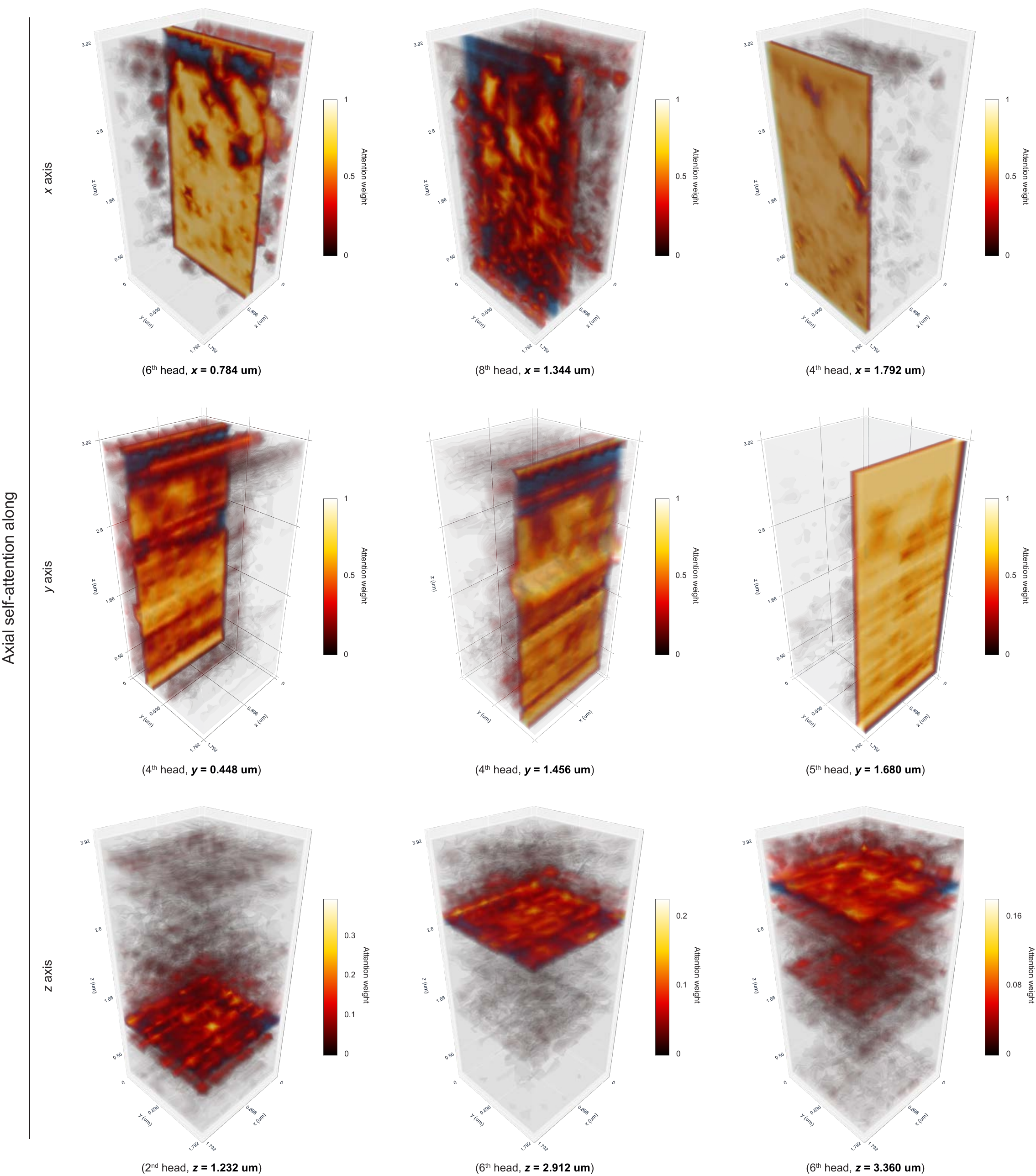}
    \caption{\textbf{Learned attention weight visualization.} Learned attention weights of multi-head axial self-attentions along each $x,y,z$ axis. Parentheses contain information on the selected attention head and the position of the layer of interest (blue, $s_i$) that attends to all layers $\left(s_j\:\left(j = 1, 2, \cdots, N\right)\right)$ with attention weights (red, $\alpha_{ij}^k$), showing importance of $s_j$ to $s_i$.}
    \label{fig:attention_visualization}
\end{figure}

The encoder's low-dimensional manifolds are further enhanced by the axial self-attention which was proposed in order to reduce the computational complexity of multi-head self-attention (``self-attention'' in short)~\cite{vaswani2017attention}. The axial self-attention factorizes 3D self-attention into three 1D axial self-attention modules, thus reducing the complexity from $O(N^3)$ to $O(3N)$. Each axial self-attention attends to voxels along one of $x, y, z$ axes. Fig.~\ref{fig:attention_visualization} visualizes learned attention weights $p_{ij}$ that quantifies normalized ``contribution'' of other layers $s_j\:(j=1,\cdots,N)$ to the layer $s_i$. We assume that the information of layer $s_i$ is spread along the layers $s_j\:(j=1,\cdots,N)$ due to lack of spatial resolution; therefore, the axial self-attention gathers the scattered information from the layers to resolve layer $s_i$ with global-range interactions. Note that in this paper, we used Pytorch instead of the original Tensorflow implementation~\cite{wang2020axial}, and our code should be publicly available in \url{https://github.com/iksungk/APT}.

\subsection*{Training and testing environments}\label{sec:train_test}

To prepare a paired dataset for training and testing, both of the Approximant and the gold standard are divided into smaller volumes of $1.792\times 1.792\times 3.92\:\mu\text{m}^3$ with $50\%$ lateral overlap. Then, we split the paired dataset into two non-overlapping sub-datasets. One set is reserved for training, and the other for testing.
The training and test samples were drawn so as to not be correlated accidentally by spatial overlap during the ptycho- and tomo-scan operations. 

For training, we use negative Pearson Correlation coefficient (NPCC) as the training loss function~\cite{li2018imaging,goy2019high,kang2021dynamical} and the Adam optimizer for stochastic gradient descent optimization~\cite{kingma2014adam} with initial learning rate of $2\times 10^{-4}$, $\beta_1 = 0.9$, $\beta_2 = 0.999$, and without weight decay. We also update the learning rate schedule according to a polynomial rule~\cite{cheng2020panoptic} as
\begin{equation}
    lr(\text{epoch}) = lr(0)\times \left(1 - \frac{\text{epoch}}{T}\right)^p,
\end{equation}
where $T = 200$ and $p = 0.9$. We run the training process for $150$ epochs and stabilize it by a mini-batch learning strategy~\cite{li2014efficient} with batch size equal to $4$. Upon completion of the training process, the network is loaded and fixed with the optimal weights, and used to reconstruct the test volume ($4.48\times 93.2\times 3.92\:\mu\text{m}^3$), as shown in Figs.~\ref{fig:qualitative_comparison_ds}, \ref{fig:qualitative_comparison_pms} and \ref{fig:psd_reconstruction_comparison}.

For all computational procedures, \textit{i.e.} pre-processing and training \& testing processes, we used the MIT Supercloud with a Intel Xeon Gold 6248 CPU with 384~GB RAM and dual NVIDIA Volta V100 GPUs with 32~GB VRAM. Once the network was trained, it took $45$ seconds to generate the reconstruction over the test volume.

\subsection*{Quantitative metrics}\label{sec:metrics}

Because each voxel on an IC is occupied by a single material, even if ICs are printed with various materials such as copper, aluminum, and tungsten, ICs can be comfortably classified into $M$-ary labels irrespective of the printing material. To further simplify, we binarize the gold standard by thresholding according to the presence of a metal or silicon within each voxel. The gold standard reconstruction, however, may still be ambiguous especially for longitudinal features due to the missing wedge in the Fourier domain as it still does not cover the entire angular range, \textit{i.e.} $\pm 90^\circ$, due to the tomographic scheme. Since the gold standard also suffers from extensive errors in these ambiguous layers, we exclude them from our quantitative evaluations as well. More details can be found in Supplementary Materials.

The quantitative comparisons in Figs.~\ref{fig:ds_bar_threshold} and~\ref{fig:pms_bar_threshold} use four different quantitative metrics to illustrate different aspects of the reconstructions. The first two are correlative metrics: the PCC~\cite{li2018imaging} and the MS-SSIM with the same weights as in the original reference~\cite{wang2003multiscale}.

The remaining two metrics are the DSC~\cite{dice1945measures} and the BER. The former is a widely accepted similarity measure in image segmentation to compare an algorithm output against its reference in medical applications~\cite{milletari2016v,zou2004statistical}. The BER measures the ratio of erroneously classified voxels over the total voxels, and it is allowable because of our binarization approach. Both of these metrics are probabilistic in the sense that they involve the estimation of probability density functions. They are obtained as
\begin{equation}\label{eq:dsc}
    \text{DSC} = \frac{2\cdot \text{TP}}{2\cdot\text{TP} + \text{FN} + \text{FP}},
\end{equation}
and
\begin{equation}\label{eq:ber}
    \text{BER} = \frac{\text{FP} + \text{FN}}{\text{TP}+\text{TN}+\text{FP}+\text{FN}},
\end{equation}
where $\text{TP}$, $\text{TN}$, $\text{FP}$, and $\text{FN}$ indicate the number of true positives, true negatives, false positives, and false negatives, respectively. For the gold standard, the binary thresholds and prior probabilities $p(0)$, $p(1)$ required for these quantities were estimated by an Expectation Maximization (EM) procedure. For the tests, we used Bayes' rule $p(x|0)p(0) = p(x|1)p(1)$ with $p(0)$, $p(1)$ same as for the gold standard.



\bibliography{scibib}

\bibliographystyle{Science}

\section*{Acknowledgments}
We are grateful to Jung Ki Song, Mo Deng, Baoliang Ge, William Harrod, Ed Cole, Zachary Levine, Bradley Alpert, Nina Weisse-Bernstein, Lee Oesterling and Antonio Orozco for helpful discussions and comments. Funding from the Intelligence Advanced Research Projects Activity, Office of the Director of National Intelligence (IARPA-ODNI), contract FA8650-17-C-9113 is gratefully acknowledged. The MIT SuperCloud and Lincoln Laboratory Supercomputing Center provided resources (high performance computing, database, consultation) that have contributed to the research results reported within this paper. I. Kang acknowledges support from Korea Foundation for Advanced Studies (KFAS). This research used resources of the Advanced Photon Source, a U.S. Department of Energy (DOE) Office of Science User Facility, operated for the DOE Office of Science by Argonne National Laboratory under Contract No. DE-AC02-06CH11357. The views and conclusions contained herein are those of the authors and should not be interpreted as necessarily representing the official policies or endorsements, either expressed or implied, of the ODNI, IARPA or the US Government.

\section*{Supplementary materials}
Figs. S1 to S5\\
Tables S1 to S2

\newpage
\newpage
\newpage
\newpage
\newpage
\newpage

\end{document}